\documentclass{aastex}
\usepackage{emulateapj5}

\slugcomment{Accepted for publication in ApJ Letter}

\shorttitle{Early-Phase Spectra of SN~2002ap}
\shortauthors{Kinugasa et al.}

\begin{document}

\title{Early-Phase Spectra of "Hypernova" SN~2002ap}


\author{K. Kinugasa\altaffilmark{1}, H. Kawakita\altaffilmark{1},
K. Ayani\altaffilmark{2}, T. Kawabata\altaffilmark{2},
H. Yamaoka\altaffilmark{3}, \\
J.S. Deng\altaffilmark{4,5},  P.A. Mazzali\altaffilmark{6,4,5},
K. Maeda\altaffilmark{5} and K. Nomoto\altaffilmark{5,4}}


\altaffiltext{1}{Gunma Astronomical Observatory, 6860-86, Nakayama,
Takayama, Agatsuma, Gunma 377-0702, Japan;
kinugasa@astron.pref.gunma.jp, kawakita@astron.pref.gunma.jp}
\altaffiltext{2}{Bisei Astronomical Observatory, 1723-70 Ohkura, Bisei,
Okayama 714-1411, Japan; ayani@bao.go.jp, kawabata@bao.go.jp}
\altaffiltext{3}{Department of Physics, Faculty of Science,
Kyushu University, Ropponmatsu, Fukuoka 810-8560, Japan;
yamaoka@rc.kyushu-u.ac.jp}
\altaffiltext{4}{Research Centre for the Early Universe, University of
Tokyo, Hongo 7-3-1, Bunkyo, Tokyo 113-0033, Japan; deng@astron.s.u-tokyo.ac.jp}
\altaffiltext{5}{Department of Astronomy, University of
Tokyo, Hongo 7-3-1, Bunkyo, Tokyo 113-0033, Japan;
maeda@astron.s.u-tokyo.ac.jp, nomoto@astron.s.u-tokyo.ac.jp}
\altaffiltext{6}{Osservatorio Astronomico, Via Tiepolo, 11, 34131
Trieste, Italy; mazzali@ts.astro.it}


\begin{abstract}

The spectral evolution of the peculiar SN~Ic 2002ap during the first 40 days 
is presented.  The spectra display very broad absorption features, which are
typical of "hypernovae".  The maximum expansion velocity measured on the
earliest spectra exceeds $ 3 \times 10^4 $ km s$^{-1}$.  The spectrum
of SN~2002ap at the epoch of maximum brightness resembles that of SN~1997ef
more than that of SN~1998bw. The spectral evolution of SN~2002ap proceeds
at about 1.5 times the rate of SN~1997ef.  The parameterized supernova
spectrum synthesis code SYNOW was used to perform line identification and 
deduce velocity information from the early-phase spectra, which are heavily
affected by line blending. The photospheric velocity, as deduced from the 
fitting results and from the blueshift of the \ion{Si}{2} $\lambda$6355 
absorption minimum, is lower than in previously studied hypernovae. 
At advanced epochs, the \ion{Si}{2} $\lambda$6355 absorption minimum
becomes difficult to distinguish. This may be caused by the growth of 
[\ion{O}{1}] $\lambda\lambda$6300, 6364 emission. Together with the
rapid spectral evolution, this suggests that SN~2002ap should enter the 
nebular phase sooner than previously studied hypernovae.
\end{abstract}


\section{Introduction}

The very powerful supernova (SN) explosions called ``hypernovae''
\citep{IWA1998} are perhaps the most energetic events in the current
universe. Their kinetic energy is $\sim 10^{52}$ ergs, which is about 
10 times larger than of normal SNe. SN~1998bw, probably associated 
with GRB 980425 \citep{GAL1998}, was the first recognized hypernova.
Although classified as a Type Ic SN (SN~Ic), the spectrum of
SN~1998bw showed unusual broad spectral features in the early phase. 
The derived velocities were as large as 30,000 km s$^{-1}$. 
The kinetic energy of SN~1998bw was deduced to be 3 -- 5 $\times$
10$^{52}$ ergs \citep{IWA1998,NAK2001}.
Other Type Ic hypernovae were subsequently recognized based on their broad
absorption features.  These include SNe~1997ef (e.g., Iwamoto et al.\
2000), 1998ey \citep{GAR1998}, and the recently discovered 2002bl
\citep{FIL2002}.  Moreover, SN~1997dq was suggested to be a hypernova from its
similarity to SN~1997ef at later phases \citep{MATH2001}.
The kinetic energy deduced for SN~1997ef was about $1.8\times10^{52}$ 
ergs (Mazzali, Iwamoto, \& Nomoto 2000), which was smaller than that of 
SN~1998bw, but still significantly larger than normal SNe. 
Hypernovae also occur among SNe~IIn, for example, SN~1997cy, 
one of the brightest SNe ever discovered and characterized by a large
kinetic energy, $\sim 3\times 10^{52}$ ergs \citep{TUR2000}, and SN~1999E
(Cappellaro, Turatto, \& Mazzali 1999). The range of properties of
hypernovae, and their possible association with GRBs are also topics of
great interest (e.g., Nomoto et al. 2001).

SN~2002ap was discovered in M74 on January 29.4 UT by Y.Hirose
\citep{NAK2002}. Spectra obtained on January 30 -- 31 
showed very broad absorption features, resembling those in SNe~1998bw 
and 1997ef. Therefore, SN~2002ap was suggested to be a hypernova 
\citep{KIN2002,MEI2002,GAL2002a}. SN~2002ap is the nearest hypernova
discovered to date, hence it was a good target even for small
telescopes. Various multi-wavelength observations of SN~2002ap have
been performed (summarized in Gal-Yam, Ofek, \& Shemmer 2002). NIR 
photometry was also performed frequently at Gunma Astronomical 
Observatory (GAO) (E.\ Nishihara et al.\ 2002, in preparation).

We report here on the optical spectroscopy of SN~2002ap performed at GAO. 
We compare the spectra of SN~2002ap with those of other hypernovae. 
We also used the parameterized
SN synthetic-spectrum code SYNOW to perform line identification and deduce
velocity information from the spectra in the early phase, when very broad
features resulting from line-blending were present.

\section{Observations and Data Reduction}

Observations of SN~2002ap started on the night of the discovery report. 
Nightly early phase spectral observations were carried out at GAO and 
Bisei Astronomical Observatory (BAO) for about 40 days, from discovery 
to solar conjunction. Spectra were obtained with Gunma Compact 
Spectrograph (GCS) attached to the classical-Cassegrain focus of the GAO
65cm telescope (F/12) and with a spectrograph attached to the BAO 1.01-m
telescope (F/12). The spectral resolving power ($\lambda/\Delta\lambda$)
of GCS is about 500 at 5500 \AA\ and the wavelength coverage is from 3800 
to 7600 \AA. The BAO spectrograph covers from 4700 to 7000 \AA\ with a
resolving power of about 1000. Observational parameters are listed in
Table 1. Since all the spectra were obtained with narrow slits (2$''$ on
GCS and 2.6$''$ at BAO, compared to a typical seeing of 2-3$''$) and
the sky conditions were generally non-photometric, the flux was not
on an absolute flux scale. The slit was aligned in E-W direction at 
GAO and in the N-S direction at BAO, which in both cases was not the 
parallactic angle. Consequently, the blue end of the spectra could be
significantly affected at advance epochs, as the observations were done 
at low altitude. In those cases, we cut the blue end of the spectra (see 
Figure 1). 

Data reduction was performed with IRAF\footnote{IRAF is distributed by 
the National Optical Astronomy Observatories, which are operated by 
the Association of Universities for Research in Astronomy, Inc., 
under cooperative agreement with the National Science Foundation.}.
Calibration for the relative intensity was performed based on the
observations of spectrophotometric standard stars. A Fe-Ne-Ar lamp for
GAO and a Fe-Ne lamp for BAO were used for the wavelength calibration.

\placetable{tbl1}

\section{Results and Discussion}

The time-sequence of the calibrated spectra of SN~2002ap is shown in
Figure 1. The evolution of the spectra is apparent. As time progresses,
most absorption features shift to the red, and the continuum becomes redder.

\begin{figure}
\epsscale{0.80}
\plotone{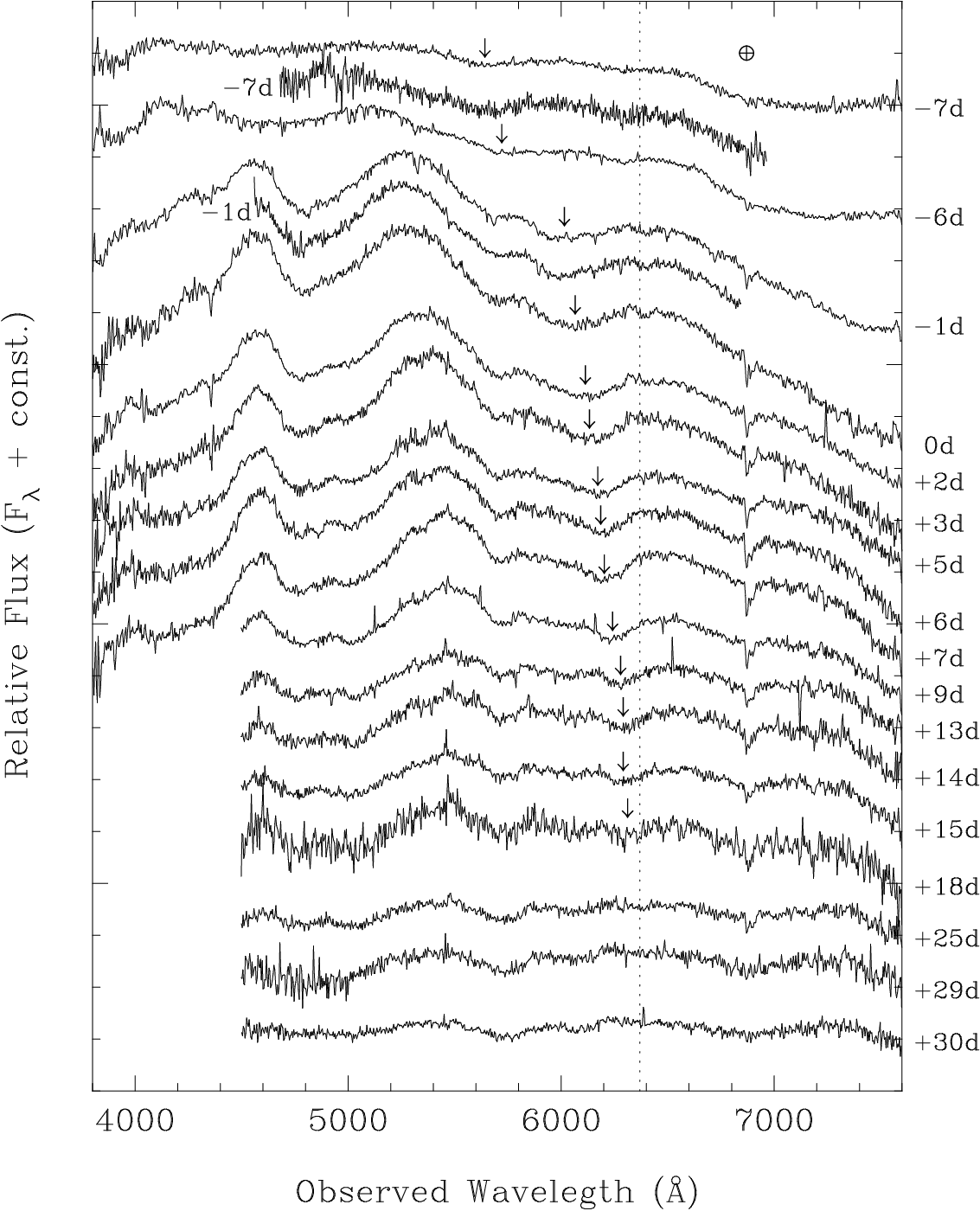}
\caption{Spectral evolution of SN~2002ap from -7 to +30 days relative
to $B$-maximum (Feb.\ 7.1; Gal-Yam et al.\ 2002). 
The spectra have been shifted vertically by arbitrary amounts. 
The horizontal axis is the observed wavelength. 
Arrows mark the absorption minima of \ion{Si}{2} $\lambda$6355 (See text 
 on the errors). The dashed line marks the rest wavelength (at the
 redshift of M74) of \ion{Si}{2}. The blue ends of  advanced-epoch
 spectra, which are affected by observing at low altitude, have been cut
 for clarity.  The symbol $\oplus$ marks the  telluric B-band
 absorption. \label{fig1}}
\end{figure}

\placefigure{fig1}

The first two spectra, taken on the same night, show a blue continuum, 
with neither deep absorptions nor strong emissions (see also the top 
of Figure 2).
Broad and shallow depressions are seen near 4700, 5700, and 6200 \AA.
These broad features resemble those of SNe~1998bw and 1997ef in the 
earliest phases \citep{PAT2001,IWA2000}.  However, SN~2002ap is bluer,
possibly because it was caught at an earlier epoch. The fact that the
first observations must have been very close to the time of explosion is
exemplified by the rapid development of spectral features. The first two 
spectra also show a broad peak near 5000\AA. But in the third 
spectrum taken the following night, this peak has shifted significantly 
to the red ($\sim 5100$\AA), indicating a rapid decrease of the material
velocity sampled by the spectra, as the photosphere moves inwards.

Photometric observation at Wise Observatory established that the peak of the
light curve occurred on February 7.1 and 8.8 UT in the $B$ and $V$-band, 
respectively \citep{GAL2002b}. We designate the epochs 
using the number of days from $B$ maximum. Figure 2 also shows the 
spectrum of SN~2002ap taken on day $+$2 (February 9.4 UT), near the time of
maximum. We compared this with the near-maximum spectra of other SNe~Ic:
SNe~1998bw, 1997ef (hypernovae) and 1994I (a normal SN~Ic;
Nomoto et al. 1994). Based on the line width, the spectra can easily
be separated into two groups. One is that of ``hypernovae'', including 
SNe~1998bw, 1997ef, and 2002ap, while the other includes only the normal
Type Ic SN~1994I. Therefore, SN~2002ap is easily identified as a
hypernova. Moreover,
the spectrum of SN~2002ap shows emission-like features around 4600 \AA\
and 5300 \AA, resembling the spectrum of SN~1997ef more than that of
SN~1998bw.

\begin{figure}
\epsscale{0.80}
\plotone{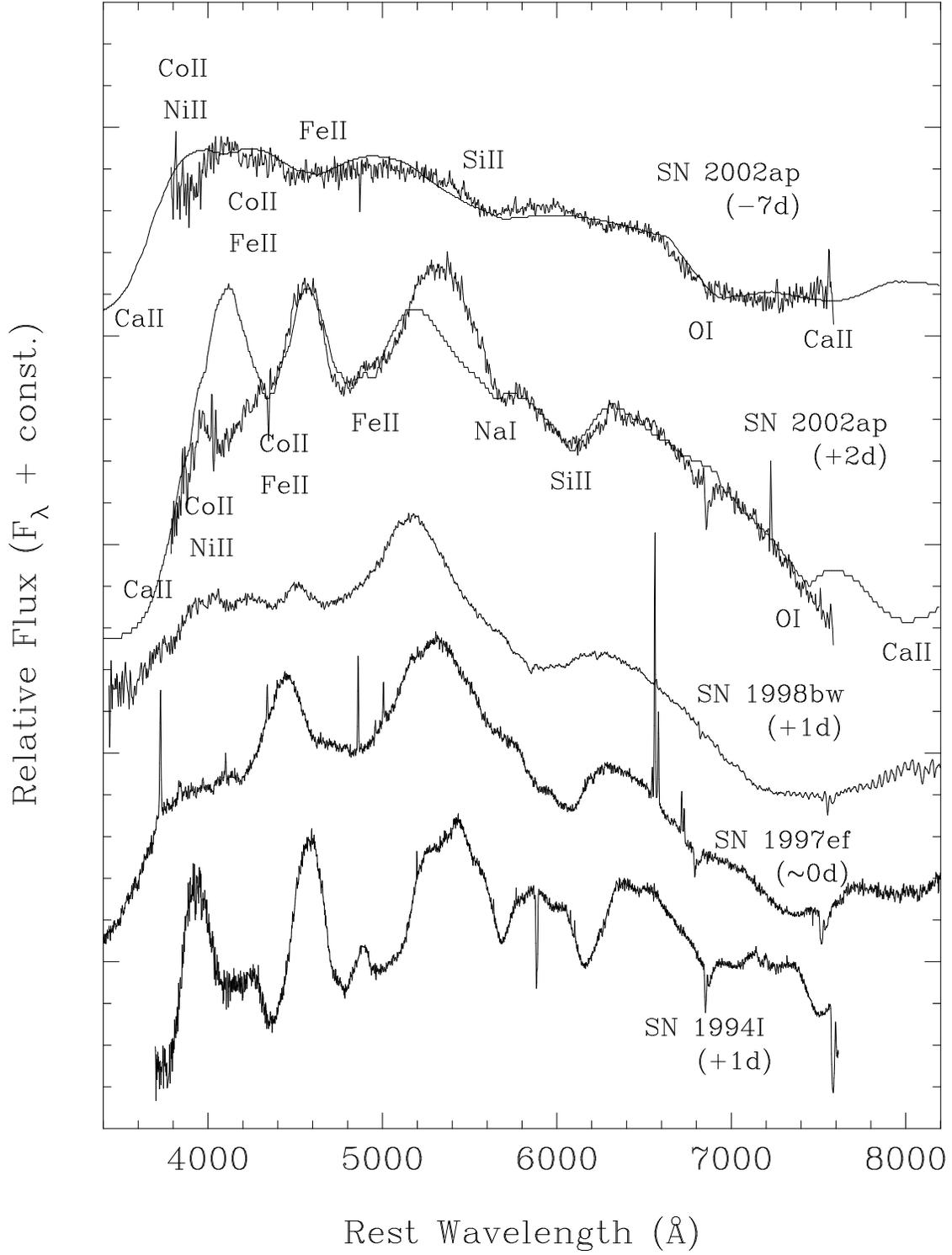}
\caption{The spectra of SN~2002ap taken on Jan.\ 31 and Feb.\ 9 are 
compared with a synthetic spectrum computed with SYNOW, and the line 
identification is shown. Also shown are the maximum-light spectra of 
other SNe~Ic for comparison (SNe~1998bw and 1997ef from Fig.1 of 
Mazzali et al.\ 2002 and SN~1994I from Fig.1b of Millard et al.\ 1999).
The spectra have been scaled and shifted vertically by arbitrary amounts, 
and shifted in wavelength to the rest-frame of the parent galaxies. 
The observed spectra of SN~2002ap have been corrected for the interstellar
extinction, $E(B-V) = 0.09$ (Takada-Hidai, Aoki, \& Zhao 2002)
and for the host galaxy recession, 657 km s$^{-1}$.
\label{fig2}}
\end{figure}

\placefigure{fig2}

In Figure 3 we compare similar-looking spectra of SNe~2002ap and 1997ef. 
We can see that the spectral evolution of SN~2002ap is about 
1.5 times faster. 
The earliest available spectrum of SN~1997ef, on day  $-$6, appears 
to have properties intermediate between the spectrum of SN~2002ap on 
day $-$1 and that on day $-$6, judging from the progressive redshifting 
of line features. The near-maximum spectrum of SN~1997ef resembles 
that of SN~2002ap on day $+$2 in all major features. 
The similarity between the spectrum of SN~1997ef on day  $+$19 and 
that of SN~2002ap on day $+$15 is striking, both showing minor features
of comparable intensities at about 4900, 5900, 6200, 7100 and 7300 \AA. 
Most of the new features visible in the spectrum of SN~1997ef on day 
$+$52, some of which are probably net emissions, are already visible 
in SN~2002ap on day $+$25, although they are not as developed. 
Both of these spectra suggest that the SNe are making a transition 
to the nebular phase, while the spectrum of SN~1997ef on day $+$27 
is still essentially photospheric in nature \citep{MAZ2000}.

\begin{figure}
\epsscale{0.80}
\plotone{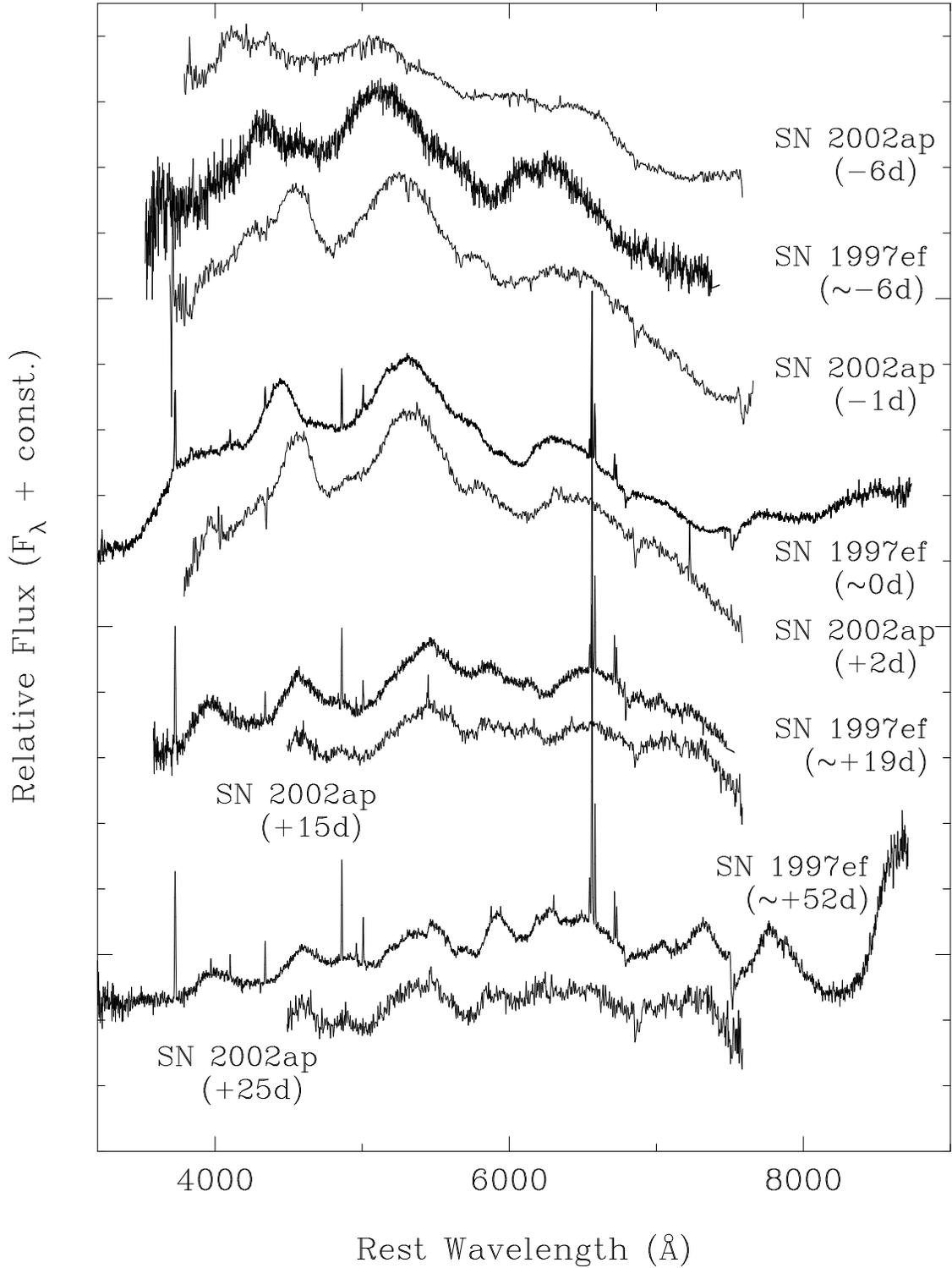}
\caption{A comparison of various spectra of SN~2002ap and
similar-looking spectra of SN~1997ef. The spectrum of SN~1997ef on 
day $\sim$ $-$6 is inserted between two adjacent SN~2002ap
 spectra. Other spectra of SN~1997ef are polluted by sharp \ion{H}{2}
 emissions and hence easily recognizable. The spectra have been scaled
 and shifted vertically by arbitrary amounts, and shifted in wavelength
 to the rest-frame of the parent galaxies. \label{fig3}}
\end{figure}

\placefigure{fig3}

Especially in the earliest spectra, most features are so broad because
of line blending that it is not easy to identify which lines contribute 
to them. Thus we used the parameterized SN spectrum synthesis code 
SYNOW, described by \citet{FIS2000}, to perform line identification and 
deduce velocity information. It is simpler and faster than the
Monte-Carlo code of \citet{MAZ2002}, and better suited for our
objective, as it does not require the  construction of realistic SN models.
In this paper, the radial-dependent line
optical depth above the photosphere is assumed to be proportional to
$(v/v_{\rm ph})^{-n}$, where $v$ is the expansion velocity and $v_{\rm
ph}$ is the photospheric velocity. 
A reasonable fit to the observations could be obtained using just a few 
ions: \ion{Ca}{2}, \ion{O}{1}, \ion{Si}{2}, \ion{Fe}{2}, \ion{Co}{2} 
and \ion{Ni}{2}, consistent with the line identification in 
SNe~1997ef and 1998bw \citep{BRA2001}.  Synthetic spectra and line 
identification for day $-$7 and $+$2 are shown in Figure 2.

Either a high $v_{\rm ph}$ or a small $n$ can make the lines broad.
Guided by Branch's (2001) work, we found 
that both are necessary to account for the extreme line widths. 
Similar results were obtained by \citet{MAZ2002}, who used a
Monte-Carlo code to synthesize some selected spectra of SN~2002ap. 
Our best fit value of the power-law index was 3
for the first spectrum.  This index value was used for the
radial dependence of the opacity for all lines and at all epochs.
As shown in Figure 4, the photospheric velocity obtained from the fitting
decrease from 35,000 $\rm km~s^{-1}$ on day $-7$ to 13,000 $\rm
km~s^{-1}$ on day $+$2, with the uncertainty being about
1000 $\rm km~s^{-1}$ empirically \citep{BRA2002}.

Our fits are reasonable for these epochs, allowing us to identify most
absorption lines.  However, some problems remain. One is the unwanted
``peak'' around 4100 \AA, which is actually the remainder of  continuum. We
can remove it by introducing both \ion{Ti}{2} and  \ion{Sc}{2}
lines. However, there is no other argument for \ion{Sc}{2} in SNe~Ic.
Such difficulties may show the limitations of SYNOW. They are probably 
due to the assumption of a pure resonant scattering source function 
and on the simple power-law dependence of the optical depth on 
velocity applied to all ions.
The hump around 5200 \AA\ is also incorrectly reproduced, probably 
the consequence of not transferring sufficient flux from the blue 
($\sim 4000$\AA) in SYNOW. 
Finally, we introduced \ion{Na}{1} to fit the absorption around 5700 \AA. 
\ion{He}{1} $\lambda$5876 may also contributes there.
The NIR spectra show a significant feature, possibly due to 
\ion{He}{1} $\lambda$10830 \citep{MOT2002}. However, that feature in
SN~1994I might have been \ion{C}{1}, \ion{O}{1}, and/or \ion{Si}{1}
lines. \citep{BAR1999,MIL1999}. 

Among those absorption lines identified, the \ion{Si}{2} $\lambda\lambda$6347,
6371 (collectively called $\lambda$6355) absorption minimum was easily
distinguishable during almost the entire observation period.
Therefore, we measured the minimum of the \ion{Si}{2} absorption as an
alternative method of estimating the photospheric velocity. In Figure 1, 
we marked with an arrow the minimum of the \ion{Si}{2} absorption. 
The evolution of the photospheric velocity, as deduced from this method, 
is also shown in Figure 4. At the earliest phases, the typical
statistical error was around $\pm$ 3000 km s$^{-1}$.
Moreover, strong line blending makes the estimate of the velocity
somewhat uncertain at those epochs. Near maximum light,
although the minima are clearly visible (the statistical error was about
$\pm$ 1500 km s$^{-1}$), the systematic error on the velocity is 
mostly due to the ambiguity introduced by the uncertain continuum. 
At late phases, the error (around $\pm$ 2500 km s$^{-1}$) was
mainly due to poor statistics.
The difference between the velocities from \ion{Si}{2} and those from
SYNOW fitting may be mostly caused by systematic errors introduced by the 
uncertainty in the continuum and the heavy line blending.

\begin{figure}
\epsscale{0.80}
\plotone{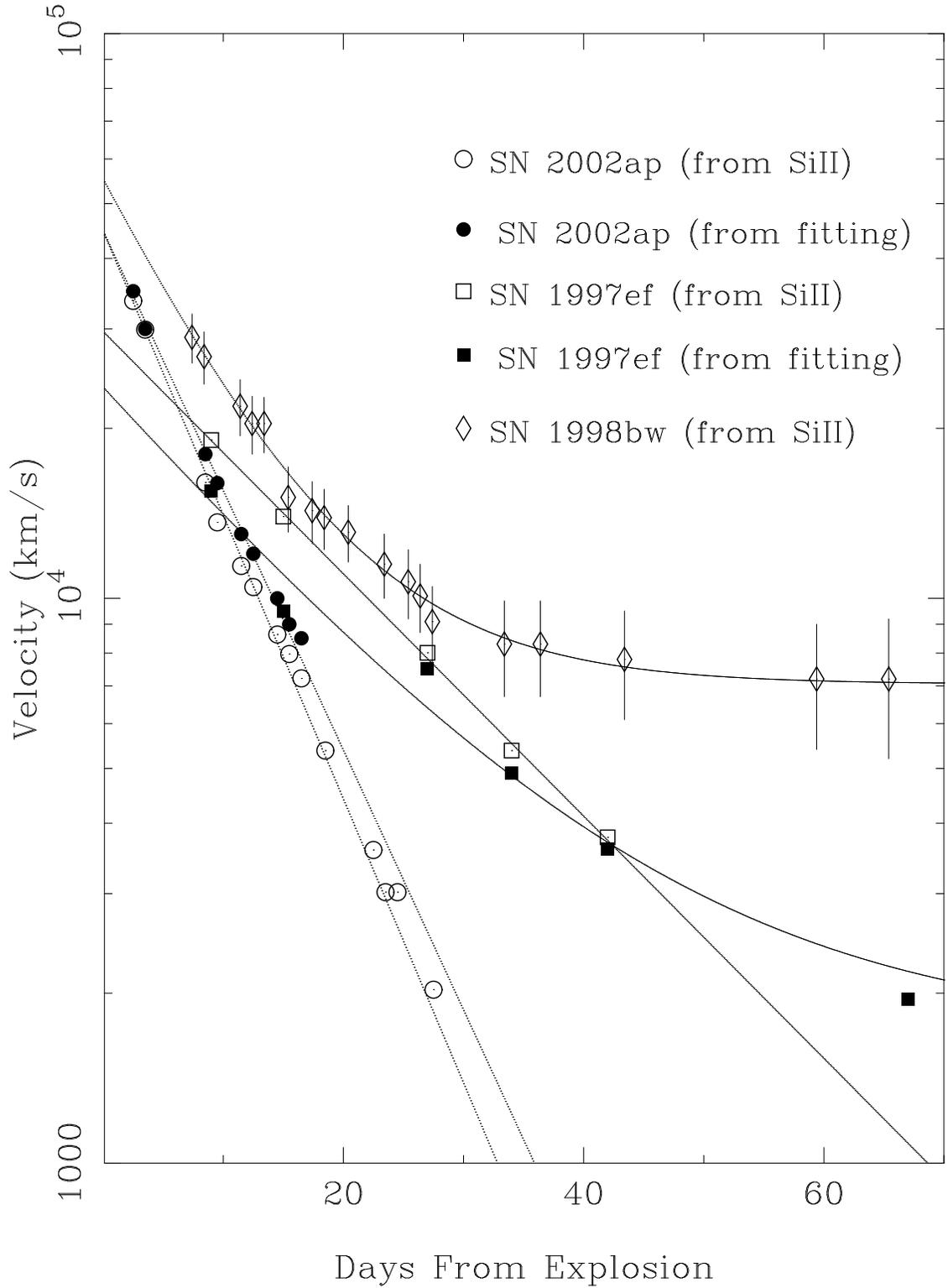}
\caption{The evolution of the photospheric velocity of hypernovae. The
SN~2002ap data are deduced from the \ion{Si}{2} $\lambda$ 6355 absorption 
minima (Fig.1) and from SYNOW fitting.  The explosion date was set at 
Jan.\ 28.9 (Mazzali et al.\ 2002). For comparison, the data for the
 other hypernovae are shown from Patat 
et al.\ (2001) for SN~1998bw, and Mazzali et al.\ (2000) for SN~1997ef.
The curves are fitted with  exponential functions for SNe~2002ap and
1997ef (\ion{Si}{2} data), and with exponential functions plus  
constants for SNe~1998bw and 1997ef (fitting data). \label{fig4}}
\end{figure}

\placefigure{fig4}

In Figure 4, we compare the evolution of the photospheric velocity of 
SN~2002ap, 1998bw and 1997ef \citep{PAT2001}. We adopted for SN~2002ap
an explosion 
date of January 28.9 UT (day $-$9.2), as proposed by \citet{MAZ2002}.
The photospheric velocity of SN~2002ap declines faster than for the
other two hypernovae, suggesting that the ejecta mass of SN~2002ap 
is smaller.
Although the velocity evolution of the three SNe is initially well fitted with
exponential functions, both SNe~1997ef and 1998bw eventually seem to flatten
out, SN~1998bw at about 7000 km s$^{-1}$ and SN~1997ef at about 2000 km
s$^{-1}$. While the measurement for SN~1998bw can be done directly on the
spectra, the value for SN~1997ef is derived from model fitting, since, as
for SN~2002ap, the \ion{Si}{2} line becomes indistinguishable at advanced
epochs (see spectra on day $+$52 for SN~1997ef and on day $+$25 for
SN~2002ap in Figure 3). No sign of the flattening for SN~2002ap was seen 
over the period of our observations, but it might occur later, which 
may have to be established through detailed spectral modeling.

The rapid evolution of the photospheric velocity and of the spectra of 
SN~2002ap suggests that SN~2002ap will develop a nebular spectrum earlier 
than the other hypernovae presented here.
In SN~1997ef nebular emission was observed in the \ion{Ca}{2} IR triplet
on day $+$27. Our spectra of SN~2002ap unfortunately do not extend
to that region.
However, the spectrum on day $+$25, which is intermediate between the
day $+$27 and the day $+$52 spectra of SN~1997ef, may show the onset of
weak [\ion{O}{1}] $\lambda\lambda$6300, 6364 net emission. 
E.\ Nishihara et al.\ (2002, in preparation) show that a break occurs in the
evolution of the IR colors around February 25, suggesting the onset of
nebular emission.
The first spectra of SN~2002ap after solar conjunction, taken on day 131 
and 140, show that it has already entered the nebular phase \citep{LEO2002}.

The broad spectral features, suggesting high kinetic energy, distinguish
hypernovae from normal SNe~Ic.
However, the observations of SN~2002ap suggest that there is a variety of
hypernovae, based on the observational features (spectra, spectral
evolution, luminosity, light curves, the association with a GRB, etc.).
More observations of hypernovae are necessary in order to understand
what physical parameters are responsible for the variations among
hypernovae and whether there is a gap in properties between hypernovae 
and normal SNe~Ic (Figure 2 of Nomoto et al.\ 2002).

%

\clearpage

\clearpage

\clearpage

\clearpage

\clearpage

\begin{deluxetable}{lcccc}
\tabletypesize{\scriptsize}
\tablecaption{Observation log. \label{tbl1}}
\tablewidth{0pt}
\tablehead{
 \colhead{Date (UT)\tablenotemark{a}} &
 \colhead{Phase (Days)\tablenotemark{b}} & \colhead{Exposure (s)} &
 \colhead{Telescope\tablenotemark{c}}
}
\startdata
     2002 Jan. 31.40 & $-$7 & 3600 & G \\
     2002 Jan. 31.47 & $-$7 & 900 & B \\
     2002 Feb. 1.40  & $-$6 & 3600 & G \\
     2002 Feb. 6.39  & $-$1 & 5400 & G \\
     2002 Feb. 6.41  & $-$1 & 2700 & B \\
     2002 Feb. 7.41  & 0 & 3600 & G \\
     2002 Feb. 9.39  & $+$2 & 3600 & G \\
     2002 Feb. 10.41  & $+$3 & 2400 & G \\
     2002 Feb. 12.39  & $+$5 & 3600 & G \\
     2002 Feb. 13.40  & $+$6 & 2400 & G \\
     2002 Feb. 14.40  & $+$7 & 3600 & G \\
     2002 Feb. 16.39  & $+$9 & 3600 & G \\
     2002 Feb. 20.39  & $+$13 & 3600 & G \\
     2002 Feb. 21.42  & $+$14 & 2400 & G \\
     2002 Feb. 22.40  & $+$15 & 3600 & G \\
     2002 Feb. 25.42  & $+$18 & 2400 & G \\
     2002 Mar. 4.41  & $+$25 & 2400 & G \\
     2002 Mar. 8.40  & $+$29 & 2400 & G \\
     2002 Mar. 9.41  & $+$30 & 1800 & G \\ \hline
\enddata

\tablenotetext{a}{Starting time of the exposures.}
\tablenotetext{b}{Days from $B$-maximum, Feb.\ 7.1 UT (MJD=52312.1)
(Gal-Yam et al.\ 2002). This was about 9.2 days after the explosion, which
 estimated by Mazzali et al.\ (2002) to have occurred on Jan.\ 28.9.}
\tablenotetext{c}{'G' and 'B' represent GAO 65cm and BAO 1.01m
 telescopes, respectively.}
\end{deluxetable}

\end{document}